\documentclass[allclo]{FBSart}
\usepackage{amsfonts}
\usepackage{amssymb}
\usepackage{epsfig}
\usepackage{graphicx}

\title{Style File for Few-Body Systems}
\author{S. Bacca \instnr{1} \thanks{\textit{E-mail address:} bacca@gsi.de}, 
N. Barnea \instnr{2} , W. Leidemann \instnr{3}, G. Orlandini  \instnr{3}
\instlist{Gesellschaft f\"ur Schwerionenforschung, Planckstr.~1, 64291
Darmstadt, Germany \and 
Racah Institute of Physics, Hebrew University, Jerusalem, Israel and
Institute of Nuclear Physics, University of Washington, Seattle, WA
98195, USA
\and Dipartimento di Fisica, Universit\`a di Trento and INFN (Gruppo
Collegato di Trento), Italy}}

\runningauthor{S.\,Bacca}
\runningtitle{Style File for Few-Body Systems}
\begin{document}
\title{ Longitudinal response function of $\mathrm {}^4 He$ with a realistic force}
\maketitle
\begin{abstract}
The longitudinal response function of  $\mathrm {}^4 He$ is calculated with
the Argonne V18 potential. 
The comparison with experiment suggests the need of a three-body force.
When adding the Urbana IX three-body potential in the calculation of the
lower longitudinal
multipoles, the total strength is suppressed in the
quasi-elastic peak, towards the trend of the
experimental data.
\end{abstract}

The inclusive electron scattering is governed by two kinds of response
functions: the
longitudinal, $R_L(\omega,{\bf q})$, and the transverse response,
$R_T(\omega,{\bf q})$, where $\omega$ and ${\bf q}$ are the energy and
momentum transfer. $R_{L/T}$ are induced by the electromagnetic charge, $\hat{\rho}({\bf q})$, and  current, $\hat{\bf J}({\bf q})$, operators, respectively.
We investigate the medium-low momentum transfer region of  the longitudinal
response function of $\mathrm {}^4 He$  starting  with a realistic
two-body interaction only, the Argonne V18 (AV18) potential \cite{AV18}, and then adding 
the Urbana IX (UIX) three-body force \cite{UIX} as a second step.
The longitudinal response is  in fact a rich soil for the investigation  of three-nucleon
force effects, since one can trust the non-relativistic limit and
one does not need to include meson exchange currents, as
required for $R_T$ to
ensure gauge invariance. 
 The longitudinal response is defined as 
\begin{equation}
R_L(\omega,{\bf q} )= \int
\!\!\!\!\!\!\!\sum_f\left| \left\langle \Psi
    _{f}\right|\hat{\rho}({\bf q})\left| \Psi
    _{0}\right\rangle \right| ^{2}\delta
\left(E_{f}-E_{0}-\omega+\frac{{\bf q}^2}{2M} \right) \,,
\label{R_L}
\end{equation}
where $E_{0/f}$ are the energies of the ground and final states
$\left|\Psi_{0/f} \right\rangle$ and $M$ is the mass of  $\mathrm {}^4 He$. In our calculation of $R_L$
the final state interaction of the continuum four-body wave function is fully taken into
account via the  Lorentz Integral Transform (LIT) method \cite{LIT}, where one does not calculate (\ref{R_L}) directly, but one first evaluates an integral transform of it.  The method
leads to the solution of a  Schr\"{o}dinger-like equation with bound
state-like asymptotics. We solve it making   use of a
 spectral resolution method based on  the construction of an 
 effective interaction  in the hyperspherical harmonics  basis (EIHH)
 \cite{EIHH, nir}. The response function itself is then obtained by a stable inversion of the integral transform. \\
\begin{figure}[t]
\vspace{70pt}
\center
\epsfig{file=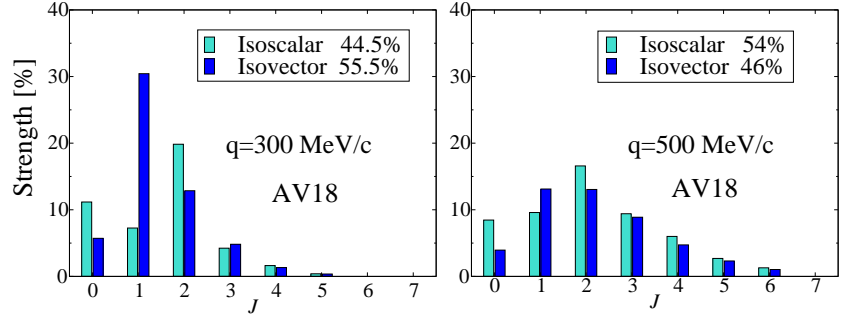, width=11cm}
\caption{Percentage contributions of multipoles of order $J$ to the
  total strength.}
\label{figure1}
\end{figure}
In the present calculation, as done in \cite{el_semirealistic},  we separate the
charge operator into isoscalar and isovector parts, which we expand in
multipoles.
In Fig.~\ref{figure1} we show the total strength of each multipole
for two different momentum transfer values, 
 ${\rm q}=300$ MeV/c and ${\rm q}=500$ MeV/c. 
One notes that the convergence of the multipole expansion gets slower with increasing momentum
transfer.
\vspace{-2.5cm}
\begin{figure}[h]
\vspace{70pt}
\center
\epsfig{file=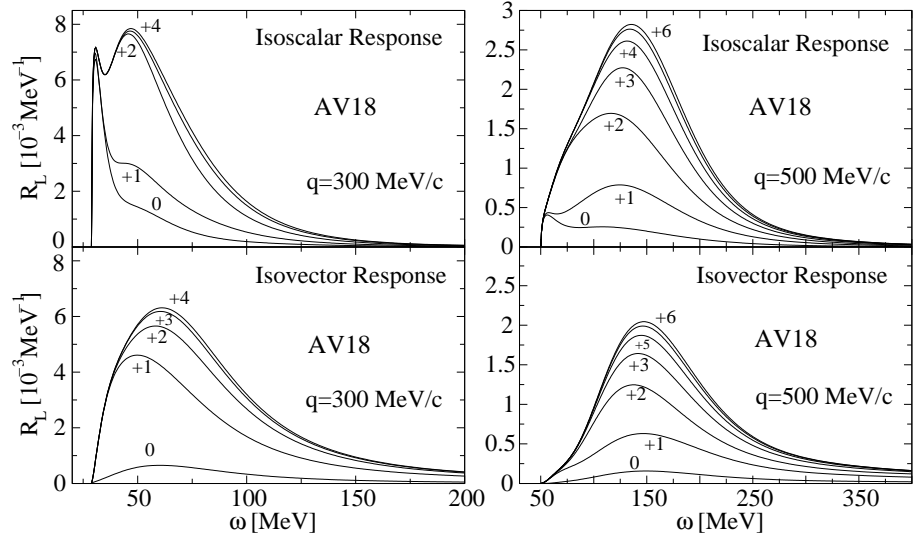, width=12cm}
\caption{Isoscalar and isovector response functions of $\mathrm {}^4 He$ for
  different multipoles, consecutively summed up to the indicated order $J$, for
${\rm q}=300$ MeV/c and ${\rm q}=500$ MeV/c.}
\label{figure2}
\end{figure}
However, with the present approach
 we can easily access the full response function of each
multipole as a function of the energy, as presented in
Fig.~\ref{figure2} for the same two momentum transfer values.
Summing then  the total isoscalar and isovector contributions, we
can directly compare our calculation with the available
experimental data.
In Fig.~\ref{figure3} one can note that the AV18 two-body potential alone
 leads to an overprediction of the quasi-elastic peak, especially for
 ${\rm q}=300$ MeV/c, suggesting the need of a three-nucleon force.
By including the UIX three-body force in the calculation
of only the monopole ($C^{S/V}_0$) and dipole ($C^{V}_1$) responses
for ${\rm q}=300$ MeV, in fact we observe a reduction of the total strength.
Since these lower multipoles, where the effect of UIX is taken into account,  contribute to about $50\%$ of the total
strength (see Fig.~\ref{figure1}), we can estimate the global effect
of the UIX force to be about twice the effect shown in Fig.~\ref{figure3}.
In future it remains to be seen
whether the AV18+UIX nuclear force model is able
to reproduce electron scattering experimental data for the
longitudinal response in the whole quasi-elastic peak region and for
different momentum transfer values.
\vspace{-2.2cm} 
\begin{figure}[htb]
\vspace{70pt}
\center
\epsfig{file=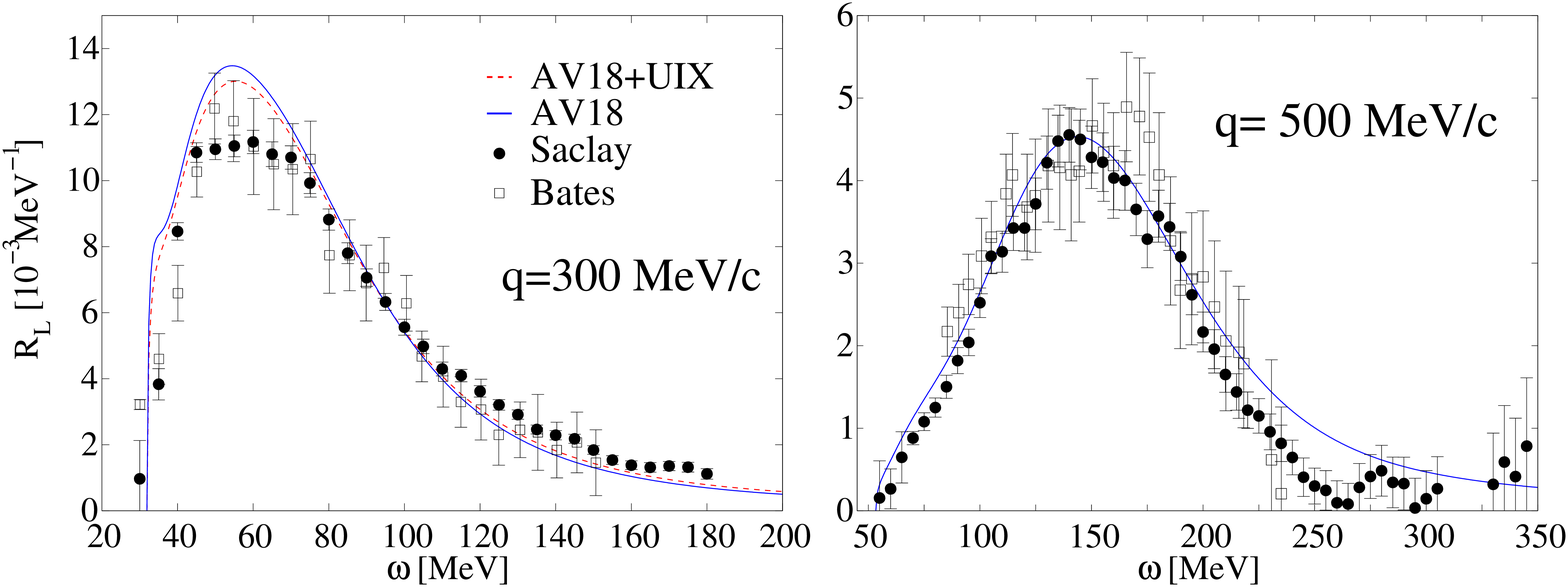, width=12cm}
\caption{Longitudinal response function of $\mathrm {}^4 He$  at momentum transfer
${\rm q}=300$ MeV/c and ${\rm q}=500$ MeV/c with the AV18 
potential in comparison with experimental data from Bates \cite{bates} and
Saclay \cite{saclay}. The curve denoted with AV18+UIX includes the three-body force
only in the  $C^{S/V}_0$ and $C^{V}_1$  multipoles.}
\label{figure3}
\end{figure}


\vspace{-0.8cm}

\begin{thebibliography}{99}
\bibitem{AV18} Wiringa, R. B., Stoks,  V. G. J.,  Schiavilla, R.: Phys. Rev. 
 {\bf C51}, 38 (1995)
\bibitem{UIX} Pudliner, B. S., et al.: Phys. Rev. {\bf C56}, 1720 (1997)
\bibitem{LIT}  Efros, V. D., Leidemann, W.,  Orlandini, G.:
 Phys. Lett. \textbf{B338}, 130 (1994)
\bibitem{EIHH} Barnea, N.,   Leidemann, W.,  Orlandini, G.: Phys. Rev.  
{\bf  C61}, 054001 (2000)
\bibitem{nir} Barnea, N.,  Novoselsky, A.: Ann. Phys. (N.Y.) {\bf 256}, 192 (1997)
\bibitem{el_semirealistic} Bacca, S.,  Arenh\"{o}vel, H. ,  Barnea, N.,   Leidemann, W., Orlandini, G.: Phys. Rev.  {\bf C76}, 014003 (2007) 
\bibitem{bates}  Dytman,  S. A., et al.:  Phys. Rev.  {\bf C38}, 800 (1988)
\bibitem{saclay} Zghiche, A.,  et al.:  Nucl. Phys.  {\bf A572}, 513 (1994)
\end{thebibliography}
\end{document}